\documentclass[12pt]{article}
  \usepackage{amsfonts}
  \usepackage{amsmath}
\usepackage{amssymb}
\usepackage{amscd}
\usepackage[dvips]{graphicx}
\begin{document}

~~
\bigskip
\bigskip
\begin{center}
%\section*
{\Large {\bf{{{Equivalent forces in Newton equation from twist
deformations and noninertial coordinate frames}}}}}
\end{center}
\bigskip
\bigskip
\bigskip
\begin{center}
{{\large ${\rm {Marcin\;Daszkiewicz}}$}}
\end{center}
\bigskip
\begin{center}
\bigskip

{ ${\rm{Institute\; of\; Theoretical\; Physics}}$}

{ ${\rm{ University\; of\; Wroclaw\; pl.\; Maxa\; Borna\; 9,\;
50-206\; Wroclaw,\; Poland}}$}

{ ${\rm{ e-mail:\; marcin@ift.uni.wroc.pl}}$}

\end{center}
\bigskip
\bigskip
\bigskip
\bigskip
\bigskip
\bigskip
\bigskip
\bigskip
\bigskip
\begin{abstract}
We compare  two ways of force terms generating in the model of nonrelativistic particle moving in the presence of constant field force $\vec{F}$.
First of them uses the twist-deformed N-enlarged Newton-Hooke quantum space-times while the second one incorporates the Coriolis  transformations of classical space. Particulary, we find the conditions for which the both treatments provide
the same force terms. The presented investigations are performed by analogy to the schemes and ideas proposed in paper \cite{vs}.
\end{abstract}
\bigskip
\bigskip
\bigskip
\bigskip
\eject

\section{{{Introduction}}}

In the last time, there appeared a lot of papers dealing with
classical and quantum  mechanics (see e.g. \cite{mech}-\cite{qmnext})
as well as with field theoretical models (see e.g. \cite{field}), in
which  the quantum space-time  plays a crucial role.
The idea to use noncommutative coordinates is quite old - it goes back to Heisenberg
and was firstly formalized by Snyder in   \cite{snyder}.  Recently, however, there were found new formal
arguments based mainly on Quantum Gravity \cite{grav1}
and String Theory models \cite{string1},
indicating that space-time at Planck scale  should be
noncommutative, i.e. it should  have a quantum nature. Besides,
 the main reason for such considerations follows from the
suggestion  that relativistic space-time symmetries should be
modified (deformed) at Planck scale, while  the classical Poincare
invariance still remains valid at
larger distances \cite{1a}, \cite{1anext}.

Currently, it is well known, that in accordance with the
Hopf-algebraic classification of all deformations of relativistic
and nonrelativistic symmetries, one can distinguish three
types of quantum spaces \cite{class1}, \cite{class2} (for details see also \cite{nnh}):\\
\\
{ \bf 1)} Canonical ($\theta^{\mu\nu}$-deformed) type of quantum space \cite{oeckl}-\cite{dasz1}
\begin{equation}
[\;{ x}_{\mu},{ x}_{\nu}\;] = i\theta_{\mu\nu}\;, \label{noncomm}
\end{equation}
\\
{ \bf 2)} Lie-algebraic modification of classical space-time \cite{dasz1}-\cite{lie1}
\begin{equation}
[\;{ x}_{\mu},{ x}_{\nu}\;] = i\theta_{\mu\nu}^{\rho}{ x}_{\rho}\;,
\label{noncomm1}
\end{equation}
and\\
\\
{ \bf 3)} Quadratic deformation of Minkowski and Galilei  spaces \cite{dasz1}, \cite{lie1}-\cite{paolo}
\begin{equation}
[\;{ x}_{\mu},{ x}_{\nu}\;] = i\theta_{\mu\nu}^{\rho\tau}{
x}_{\rho}{ x}_{\tau}\;, \label{noncomm2}
\end{equation}
with coefficients $\theta_{\mu\nu}$, $\theta_{\mu\nu}^{\rho}$ and  $\theta_{\mu\nu}^{\rho\tau}$ being constants.\\
\\
Besides, it has been demonstrated in \cite{nnh}, that in the case of
so-called N-enlarged Newton-Hooke Hopf algebras
$\,{\mathcal U}^{(N)}_0({ NH}_{\pm})$ the twist deformation
provides the new  space-time noncommutativity of the
form\footnote{$x_0 = ct$.},\footnote{ The discussed space-times have been  defined as the quantum
representation spaces, so-called Hopf modules (see e.g. \cite{oeckl}, \cite{chi}), for quantum N-enlarged
Newton-Hooke Hopf algebras.}
\begin{equation}
{ \bf 4)}\;\;\;\;\;\;\;\;\;[\;t,{ x}_{i}\;] = 0\;\;\;,\;\;\; [\;{ x}_{i},{ x}_{j}\;] = %\alpha
if_{\pm}\left(\frac{t}{\tau}\right)\theta_{ij}(x)%\;\;;\;\;\alpha -
%{\rm deformation \;parameter}
\;, \label{nhspace}
\end{equation}
with time-dependent  functions
$$f_+\left(\frac{t}{\tau}\right) =
f\left(\sinh\left(\frac{t}{\tau}\right),\cosh\left(\frac{t}{\tau}\right)\right)\;\;\;,\;\;\;
f_-\left(\frac{t}{\tau}\right) =
f\left(\sin\left(\frac{t}{\tau}\right),\cos\left(\frac{t}{\tau}\right)\right)\;,$$
$\theta_{ij}(x) \sim \theta_{ij} = {\rm const}$ or
$\theta_{ij}(x) \sim \theta_{ij}^{k}x_k$ and  $\tau$ denoting the time scale parameter
 -  the cosmological constant. It should be also noted that different relations  between all mentioned above quantum spaces ({\bf 1)}, { \bf 2)}, { \bf 3)}
and { \bf 4)}) have been summarized in paper \cite{nnh}.

Recently, in the series of articles \cite{romero}-\cite{giri}, there has been discussed the impact of different (mentioned above) kinds of
quantum spaces on the dynamical structure of physical systems. Particulary, it has been demonstrated, that
in the case of classical oscillator model \cite{oscylator} as well as in the case of nonrelativistic particle moving in constant
external field force $\vec{F}$ \cite{daszwal}, there are generated by space-time noncommutativity additional force terms. Such a type
of investigation has been  performed for quantum oscillator model as well \cite{oscylator}, i.e. it was demonstrated that the  quantum space in nontrivial way affects on the spectrum of energy operator. Besides, in article \cite{toporzelek} there has been considered model of particle moving on the $\kappa$-Galilei space-time in the presence of gravitational field force. It has been demonstrated, that in such a case
there is produced force term, which can be identified with so-called Pioneer anomaly \cite{pioneer}, and the value of deformation parameter $\kappa$ can be fixed by comparison of obtained result with observational data. It should be also noted, that from the abovementioned point of view, the case of two spatial directions commuting to the function of classical time seems to be most
interesting. As it was demonstrated in articles \cite{oscylator} and \cite{daszwal}, just this type of space-time noncommutativity produces additional time-dependent force terms, which appear in Hamiltonian function of the models. Usually, such a situation is interpreted as the interaction (by  radiation)  of considered system with (some) external source \cite{24}, \cite{24a}.

In this article we generate the (additional) time-dependent force terms in Newton equation of particle moving in constant force $\vec{F}$ in the completely new way, i.e. by the   transformation of classical space of the form
\begin{eqnarray}
&x_i& \longrightarrow \;\;x_i+ a_i(t)\;,\label{trans1}\\
&t& \longrightarrow \;\;t\;, \label{trans2}
\end{eqnarray}
with arbitrary real time-dependent function $a_i(t)$. Formally, the rules (\ref{trans1}) and (\ref{trans2}) can be realized in the framework of so-called infinite-dimensional  (classical) Coriolis group, for which \cite{coriolis} (see also \cite{coriolis2})\footnote{The Coriolis group is the biggest group of "isometries" of "flat" Galilei space-time.}
\begin{eqnarray}
a_i(t) &=& \sum_{n=0}^{\infty}a_{in} t^n\;,\label{realization}
\end{eqnarray}
with $a_{in}$ denoting the parameters of space-time transformation. Further, we compare the obtained in such a way results with the mentioned above model  of nonrelativistic particle defined
 on the  following quantum space\footnote{We consider particle moving in the presence of external constant field force $\vec{F}$.}
  %\cite{Daszkiewicz:2011jq}
\begin{equation}
[\;t,\bar{x}_{i}\;] =[\;\bar{x}_{1},\bar{x}_{3}\;]  = [\;\bar{x}_{2},\bar{x}_{3}\;]  =
0\;\;\;,\;\;\; [\;\bar{x}_{1},\bar{x}_{2}\;]  =
if({t})\;\;;\;\;i=1,2,3
\;. \label{spaces}
\end{equation}
It should be noted that the most general form of function $f(t)$ can be get  in the framework of twist procedure of N-enlarged Newton-Hooke
Hopf algebra $\;{\mathcal U}^{(N)}_{0}({{NH}}_{\pm})$ \cite{nnh}. Particularly, the explicit realization of such spaces has been provided for 6-enlarged Newton-Hooke quantum group in article \cite{nnh}. Here, we analyze the case of (six) most complicated 6-enlarged Newton-Hooke space-times  given by
\begin{eqnarray}
f({t})&=&
f_{\kappa_1}\left(\frac{t}{\tau}\right) =
518400\kappa_1\,\tau^{12}\left(\pm C_{\pm} \left(\frac{t}{\tau}\right)  \mp \frac{1}{24}\left(\frac{t}{\tau}\right)^4 - \frac{1}{2}
\left(\frac{t}{\tau}\right)^2 \mp 1\right)^2\;, \label{w2}\\
f({t})&=&
f_{\kappa_2}\left(\frac{t}{\tau}\right) = 86400\kappa_2\,\tau^{11}
\left(\pm C_{\pm} \left(\frac{t}{\tau}\right)  \mp \frac{1}{24}\left(\frac{t}{\tau}\right)^4 - \frac{1}{2}
\left(\frac{t}{\tau}\right)^2 \mp 1\right) \,\times \nonumber\\
&~~&~~~~~~~~~~~~~~~~\times~\;\left(S_{\pm} \left(\frac{t}{\tau}\right)  \mp \frac{1}{6}\left(\frac{t}{\tau}\right)^3 - \frac{t}{\tau}\right)\;,
\label{w3}\\
f({t})&=&
f_{\kappa_3}\left(\frac{t}{\tau}\right) = 17280\kappa_3\,
\tau^{10}
\left(\pm C_{\pm} \left(\frac{t}{\tau}\right)  \mp \frac{1}{24}\left(\frac{t}{\tau}\right)^4 - \frac{1}{2}
\left(\frac{t}{\tau}\right)^2 \mp 1\right) \,\times \nonumber\\
&~~&~~~~~~~~~~~~~~~~\times~\;\left(C_{\pm} \left(\frac{t}{\tau}\right)  \mp \frac{1}{2}\left(\frac{t}{\tau}\right)^2 - 1\right)
 \;, \label{w4}\\
f({t})&=&
 f_{\kappa_4}\left(\frac{t}{\tau}\right) = \pm4320\kappa_4\,\tau^{9}
\left(\pm C_{\pm} \left(\frac{t}{\tau}\right)  \mp \frac{1}{24}\left(\frac{t}{\tau}\right)^4 - \frac{1}{2}
\left(\frac{t}{\tau}\right)^2 \mp 1\right) \,\times \nonumber\\
 &~~&~~~~~~~~~~~~~~~~\times~\;\left(S_{\pm} \left(\frac{t}{\tau}\right)  - \frac{t}{\tau}\right)
\;, \label{w5}\\
f({t})&=&
f_{\kappa_5}\left(\frac{t}{\tau}\right) =  \pm1440\kappa_5\,\tau^8
\left(\pm C_{\pm} \left(\frac{t}{\tau}\right)  \mp \frac{1}{24}\left(\frac{t}{\tau}\right)^4 - \frac{1}{2}
\left(\frac{t}{\tau}\right)^2 \mp 1\right) \,\times \nonumber\\
&~~&~~~~~~~~~~~~~~~~\times~\;\left(C_{\pm} \left(\frac{t}{\tau}\right)  - 1\right)
\;, \label{w6}\\
f({t})&=&
f_{\kappa_6}\left(\frac{t}{\tau}\right) = 720\kappa_6\,\tau^7
\left(\pm C_{\pm} \left(\frac{t}{\tau}\right)  \mp \frac{1}{24}\left(\frac{t}{\tau}\right)^4 - \frac{1}{2}
\left(\frac{t}{\tau}\right)^2 \mp 1\right)
\,\times \nonumber\\
&~~&~~~~~~~~~~~~~~~~\times~\;S_{\pm} \left( \frac{t}{\tau}\right)
\;,\label{w7}
\end{eqnarray}
as well as (for $\tau$ approaching  infinity) by
\begin{eqnarray}
f(t) &=& f_{\kappa_1}({t}) = \kappa_1\,t^{12}\;,\label{ggnw2}\\
f(t) &=& f_{\kappa_2}({t}) = \kappa_2\,t^{11}\;,\label{ggnw3}\\
f(t) &=& f_{\kappa_3}({t}) = \kappa_3\,t^{10}\;,\label{ggnw4}\\
f(t) &=& f_{\kappa_4}({t}) = \kappa_4\,t^9\;,  \label{ggnw5}\\
f(t) &=& f_{\kappa_5}({t}) = \kappa_5\,t^8\;, \label{ggnw6}\\
f(t) &=& f_{\kappa_6}({t}) = \kappa_6\,t^7\;, \label{ggnw7}
\end{eqnarray}
with
$$C_{+/-} \left(\frac{t}{\tau}\right) = \cosh/\cos \left(\frac{t}{\tau}\right)\;\;\;{\rm and}\;\;\;
S_{+/-} \left(\frac{t}{\tau}\right) = \sinh/\sin
\left(\frac{t}{\tau}\right) \;.$$
Consequently, by the comparison of  both treatments we find the direct link
between transformation functions $a_i(t)$ and  time-dependent noncommutativity $f(t)$. More preciously, we write the transformation coefficients $a_{in}$  in terms of deformation parameters $\kappa_a$ (see formula (\ref{equal1bbb})) in such a way, that the force terms generated in both
approaches become the same.

Finally, it should be noted, that we use in the present article the methods and ideas already proposed in paper \cite{vs}. However, it should be also   mentioned, that the performed studies concern the most general (known) mathematical  objects such as the Coriolis group as well as the twist-deformed N-enlarged Newton-Hooke Hopf algebra. For this reason the presented investigations give (in the opinion of author) the new and original contribution  to the  noncommutative geometry area.

The paper is organized as follows. In second section  we consider
the standard (commutative) model of nonrelativistic particle moving in constant field force $\vec{F}$, which incorporates the transformation rules (\ref{trans1}) and  (\ref{trans2}). Further, we provide its noncommutative counterpart defined on quantum space (\ref{spaces}). In section three we compare the both (described above) force term generating  procedures. The final remarks are mentioned in the last section.

\section{{{The first treatment: Generating of the additional force terms in Newton equation by the Coriolis transformation of classical space (\ref{trans1})-(\ref{realization})}}}

Let us consider  the  model in which the force terms are generated by the transformation of classical space (\ref{realization}). Firstly, we start
with the following equation of motion $(\dot{{o}}_i = \frac{d}{dt}o_i)$
\begin{eqnarray}
\left\{\begin{array}{rcl} m\ddot{x}_1  &=&{F_1} \\
 &~~&~\cr
 m\ddot{x}_2  &=&
{F_2} \\
 &~~&~\cr
 m\ddot{x}_3  &=& {F_3}
 \;,\end{array}\right.\label{start}
\end{eqnarray}
defined on the commutative  space-time.
Next, by
using transformation rules (\ref{trans1}) with function $a_3(t)$ equal zero, we get the following Newton law in the nonrelativistic space-time with
changed space coordinates (see (\ref{trans1}))
\begin{eqnarray}
\left\{\begin{array}{rcl} m\ddot{x}_1  &=&{F_1} + m{\ddot a}_1(t) = G_1(t)\\
 &~~&~\cr
 m\ddot{x}_2  &=&
{F_2} + m\ddot{a}_2(t) = G_2(t)\\
 &~~&~\cr
 m\ddot{x}_3  &=& {F_3} = G_3
 \;,\end{array}\right.\label{secondnewton1}
\end{eqnarray}
i.e. there appeared in  Newton equation the additional force term given by the function ${\ddot a}_i(t)$. Moreover,
 one should  observe that the solution of (\ref{secondnewton1}) is given by
 \begin{eqnarray}
\left\{\begin{array}{rcl} {x}_{1}(t) &=& \frac{F_1}{2m}t^2  + v^0_1t + x^0_1 + a_1(t)\\
 &~~&~\cr
{x}_{2}(t) &=& \frac{F_2}{2m}t^2  + v^0_2t + x^0_2 +a_2(t)\\
 &~~&~\cr
{x}_{3}(t) &=& \frac{1}{2m}F_3t^2  + v^0_3t +x^0_3\;,\end{array}\right.
\label{secondsol3}
\end{eqnarray}
with $v^0_i$ and $x_i^0$ denoting the initial velocity and position of moving particle; in terms of coefficients $a_{in}$  it takes the form
\begin{eqnarray}
\left\{\begin{array}{rcl} {x}_{1}(t) &=& \sum_{n=3}^{\infty} a_{1n}t^n+\left(\frac{F_1}{2m}+ a_{12}\right)t^2  + \left(v^0_1+a_{11}\right)t +
x^0_1+a_{10}   \\
 &~~&~\cr
{x}_{2}(t) &=& \sum_{n=3}^{\infty}a_{2n}t^n +\left(\frac{F_2}{2m}+a_{22}\right)t^2  + \left(v^0_2+a_{21}\right)t + x^0_2+a_{20} \\
 &~~&~\cr
{x}_{3}(t) &=& \frac{1}{2m}F_3t^2  + v^0_2t +x^0_3\;.\end{array}\right.
\label{secondsolform}
\end{eqnarray}
Consequently, we see that  there appeared in model the new,
time-dependent force $\vec{G}(t) = \left[\;G_1(t),G_2(t),G_3\;\right]$
defined by the equation (\ref{secondnewton1}).

\section{{{The second treatment: Generating of the additional force terms in Newton equation by the space-time noncommutativity (\ref{spaces})}}}

Let us now turn to the second dynamical models in which the additional force terms are generated by  space-time noncommutativity. Firstly, we start with the
following phase space\footnote{We use the correspondence relation $\{\;a,b\;\}
= \frac{1}{i}[\;\hat{a},\hat{b}\;]$  $(\hbar = 1)$.}
\begin{equation}
\{\;t,{ {\bar x}}_{i}\;\} = 0 \;\;\;,\;\;\;\{\;{ {\bar x}}_{1},{ {\bar
x}}_{2}\;\}  = f(t)\;\;\;,\;\;\;
\{\;{ {\bar x}}_{1},{{\bar
x}}_{3}\;\}= 0=\{\;{ {\bar x}}_{2},{ {\bar
x}}_{3}\;\} \;, \label{beyond}
\end{equation}
\begin{equation}
\{\;{ {\bar x}}_{i},{\bar p}_j\;\} = \delta_{ij}\;\;\;,\;\;\;\{\;{
{\bar p}}_{i},{ {\bar p}}_{j}\;\} = 0\;, \label{genin2a}
\end{equation}
corresponding to the commutation relations (\ref{spaces}). One can check that the  relations (\ref{beyond}),
(\ref{genin2a}) satisfy the Jacobi identity and for deformation
parameters $\kappa_a$ running to zero become classical. Next, we define
the Hamiltonian function for
nonrelativistic particle moving in constant field force $\vec{F}$ as follows %\cite{Daszkiewicz:2011jq}
\begin{eqnarray}
H(\bar{p},\bar{x}) = \frac{1}{2m}\left({\bar p}_{1}^2 +
{\bar p}_{2}^2 + {\bar p}_{3}^2\right) -  \sum_{i=1}^{3}F_i \bar{x}_i\;.\label{ham1}
\end{eqnarray}
In order to analyze the above system we represent the
noncommutative variables $({\bar x}_i, {\bar p}_i)$ on classical
phase space $({ x}_i, { p}_i)$ as  (see e.g. \cite{giri})
\begin{equation}
{\bar x}_{1} = { x}_{1} - \frac{f_{\kappa_a}(t)}{2}
p_2\;\;\;,\;\;\;{\bar x}_{2} = { x}_{2} +\frac{f_{\kappa_a}(t)}{2}
p_1\;\;\;,\;\;\; {\bar x}_{3}= x_3 \;\;\;,\;\;\; {\bar p}_{i}=
p_i\;, \label{rep}
\end{equation}
where
\begin{equation}
\{\;x_i,x_j\;\} = 0 =\{\;p_i,p_j\;\}\;\;\;,\;\;\; \{\;x_i,p_j\;\}
=\delta_{ij}\;. \label{classpoisson}
\end{equation}
Then, the  Hamiltonian (\ref{ham1})  takes the form
\begin{eqnarray}
{{H}}({ p},{ x})=H_f(t) = \frac{1}{2m}\left({ p}_{1}^2 +
{ p}_{2}^2 + { p}_{3}^2\right) - \sum_{i=1}^{3}F_i x_i + F_1 \frac{f_{\kappa_a}(t)}{2}p_2 -
F_2 \frac{f_{\kappa_a}(t)}{2}p_1 \label{hamoscnew}\;,
\end{eqnarray}
while the corresponding  Newton equation look as follows
\begin{equation}
\left\{\begin{array}{rcl} m\ddot{x}_1  &=&{F_1} - \frac{m\dot{f}_{\kappa_a}(t)
}{2}F_2 = H_1(t)\\
 &~~&~\cr
 m\ddot{x}_2  &=&
{F_2} + \frac{m\dot{f}_{\kappa_a}(t)
}{2}F_1 = H_2(t)\\
 &~~&~\cr
 m\ddot{x}_3  &=& {F_3} = H_3
 \;.\end{array}\right.\label{dddmixednewton1}
\end{equation}

%First of all,
Firstly, by trivial integration one can find the following  solution of the above system
 %(\ref{dddmixednewton1})
 \begin{eqnarray}
 \left\{\begin{array}{rcl}{x}_{1}(t) &=& \frac{F_1}{2m}t^2  + v^0_1t - \frac{F_2}{2}\int_{0}^{t}{f}_{\kappa_a}(t')dt'\\
 &~~&~\cr
 {x}_{2}(t) &=& \frac{F_2}{2m}t^2  + v^0_2t + \frac{F_1}{2}\int_{0}^{t}{f}_{\kappa_a}(t')dt'\\
 &~~&~\cr
{x}_{3}(t) &=& \frac{1}{2m}F_3t^2  + v^0_2t +x^0_3\;,\end{array}\right.\label{sol3}
\end{eqnarray}
which in the case of functions (\ref{w2})-(\ref{w7}) is given by
\begin{eqnarray}
 \left\{\begin{array}{rcl}{x}_{1}(t) &=& \frac{F_1}{2m}t^2  + v^0_1t - \frac{F_2}{2}b(t)\\
 &~~&~\cr
 {x}_{2}(t) &=& \frac{F_2}{2m}t^2  + v^0_2t + \frac{F_1}{2}b(t)\\
 &~~&~\cr
{x}_{3}(t) &=& \frac{1}{2m}F_3t^2  + v^0_2t +x^0_3\;,\end{array}\right.\label{kksol3}
\end{eqnarray}
with
\begin{eqnarray}
b(t) &=& b_{\kappa_1}\left(\frac{t}{\tau}\right) =  900{\kappa_1}\,{\tau^4} \left(144\tau^9S_{\pm}\left(\frac{2t}{\tau}\right)+192\tau^6 t (12\tau^2 \pm t^2)C_{\pm}\left(\frac{t}{\tau}\right)+ \right. \label{bggnw2} \\
&~&~~~~~~~-\left. 48\tau^5 (72\tau^4 \pm 24\tau^2t^2 +t^4)S_{\pm}\left(\frac{t}{\tau}\right) \right. + \nonumber \\
&~&~~~~~~~+\left. \frac{1}{315}t(272160\tau^8 \pm 60480 \tau^6t^2 +12096 \tau^4t^4 \pm 1080\tau^2t^6+35t^8)\right)
\;,\nonumber\\
b(t) &=& b_{\kappa_2}\left(\frac{t}{\tau}\right) = 75\kappa_2\,\tau^4 \left(-24 \tau^4 C_{\pm}\left(\frac{t}{\tau}\right) +24\tau^4 \pm12 \tau^2t^2 +t^4
\right)^2\;,\label{bggnw3}
\end{eqnarray}
\begin{eqnarray}
b(t) &=& b_{\kappa_3}\left(\frac{t}{\tau}\right) = \frac{72}{7}\kappa_3\,\tau^4\left(\pm 420\tau^7S_{\pm}\left(\frac{2t}{\tau}\right)\pm 2520\tau^6t + 560\tau^4t^3
+ \right. \nonumber\\
&~&~~~~~~~\pm \left. 98 \tau^4t(18\tau^2 \pm t^2)C_{\pm}\left(\frac{t}{\tau}\right) -70\tau^3(120\tau^4 \pm 36\tau^2t^2 +t^4)S_{\pm}\left(\frac{t}{\tau}\right)
+\right. \label{bggnw4}\\
&~&~~~~~~~+ \left. 5t^7\right)\;,\nonumber\\
b(t) &=& b_{\kappa_4}\left(\frac{t}{\tau}\right) = 180\kappa_4\,\tau^4 \left(\pm6\tau^6 C_{\pm}\left(\frac{2t}{\tau}\right) \pm 42\tau^6\mp\tau^2(48\tau^4\pm 24\tau^2t^2+t^4)
\right. \times \nonumber \\
&~&~~~~~~~\times \left. C_{\pm}\left(\frac{t}{\tau}\right) +\frac{1}{6}t(t^2 \pm 6\tau^2)(\pm 24\tau^3S_{\pm}\left(\frac{t}{\tau}\right) \pm12\tau^2t + t^3)\right)
\;,  \label{bggnw5}\\
b(t) &=& b_{\kappa_5}\left(\frac{t}{\tau}\right) = 12\kappa_5\,\tau^4 \left(30\tau^5S_{\pm}\left(\frac{2t}{\tau}\right)+180\tau^4t\mp20\tau^2t^3+20\tau^2t(12\tau^2\mp t^2)
\right. \times \nonumber \\
&~&~~~~~~~\times \left. C_{\pm}\left(\frac{t}{\tau}\right) -5\tau(96\tau^4\mp 24\tau^2t^2 + t^4)S_{\pm}\left(\frac{t}{\tau}\right)+t^5\right)\;,
\label{bggnw6}\\
b(t) &=& b_{\kappa_6}\left(\frac{t}{\tau}\right) = 30\kappa_6\,\tau^4 \left( 6\tau^4 \left(C_{\pm}\left(\frac{2t}{\tau}\right)+11\right) +4\tau t(t^2\pm 12\tau^2)S_{\pm}\left(\frac{t}{\tau}\right)\right. + \nonumber \\
&~&~~~~~~~- \left.(72\tau^4\pm 24\tau^2t^2+t^4)C_{\pm}\left(\frac{t}{\tau}\right)\right)\;,
\label{bggnw7}
\end{eqnarray}
respectively. Secondly, one should  observe that as in the pervious approach  the noncommutativity (\ref{spaces}) generates the new, time-dependent force term
$\vec{H}(t) = \left[\;H_1(t),H_2(t),H_3\;\right]$, which for deformation parameters $\kappa_a$ approaching
zero reproduces undeformed force $\vec{F}$.

\section{Comparison  of the both approaches}

Let us now compare the formulated above treatments.
Firstly, one can observe that the new, time-dependent forces $\vec{G}(t)$ and $\vec{H}(t)$ are exactly the same when
\begin{eqnarray}
\ddot{a}_1(t) = - \frac{\dot{f}_{\kappa_a}(t)
}{2}F_2\;\;\;,\;\;\;
\ddot{a}_2(t) =  \frac{\dot{f}_{\kappa_a}(t)
}{2}F_1 \;.\label{triondnewpot1}
\end{eqnarray}
Further, it should be noted that if one expands the functions $f_{\kappa_a}(t)$ in power series with respect time variable
\begin{eqnarray}
f_{\kappa_a}(t) = \sum_{n=0}^{\infty} \frac{1}{n!} \left. \frac{d^n f_{\kappa_a}(t)}{dt^n}\right|_{t=0} t^n\;,\label{codi1b}
\end{eqnarray}
then,  the above condition can be written as follows
\begin{eqnarray}
-\sum_{n=2}^{\infty} n(n-1)a_{1n} t^{n-2} &=& \frac{F_2}{2}\left[\;\sum_{n=1}^{\infty}\frac{1}{(n-1)!} \left. \frac{d^n f_{\kappa_a}(t)}{dt^n}\right|_{t=0} t^{n-1}\;\right]\;,\label{codi2}\\
\sum_{n=2}^{\infty} n(n-1)a_{2n} t^{n-2} &=& \frac{F_1}{2}\left[\;\sum_{n=1}^{\infty}\frac{1}{(n-1)!} \left. \frac{d^n f_{\kappa_a}(t)}{dt^n}\right|_{t=0} t^{n-1}\;\right]\;.\label{codddi2}
\end{eqnarray}
Consequently, we get the following link between coefficients of classical Coriolis transformation $a_{in}$  and the deformation parameters $\kappa_a$
\begin{eqnarray}
a_{1n} = -\frac{F_2}{2n!} \left. \frac{d^{n-1} f_{\kappa_a}(t)}{dt^{n-1}}\right|_{t=0}\;\;\;,\;\;\;
a_{2n} = \frac{F_1}{2n!} \left. \frac{d^{n-1} f_{\kappa_a}(t)}{dt^{n-1}}\right|_{t=0}\;\;\;;\;\;\;n>1\;.\label{equal1bbb}
\end{eqnarray}

Next, it seems that the  treatment realized by the classical groups has one advantage - it follows from
 (\ref{secondnewton1}) and (\ref{dddmixednewton1}) that this  approach  does not need the presence
of initial constant force $\vec{F} = \left[\;F_1,F_2,F_3\;\right]$ to generate the additional dynamical terms. In fact, when
one puts $\vec{F} = \left[\;0,0,0\;\right]$ in the equation of motion (\ref{dddmixednewton1}) then its right side vanishes, while  in the equation (\ref{secondnewton1}) there are still presented the additional forces of the form
\begin{eqnarray}
G_1 (t) = m\ddot{a}_1(t)\;\;\;,\;\;\;G_2 (t) = m\ddot{a}_2(t)\;.\label{still}
\end{eqnarray}

Finally, one should to notice that the first treatment is much more general than the second one, i.e. the  coefficients of classical transformation $a_{in}$
are real and arbitrary while the functions $f_{\kappa_a}(t)$ are given by the particular realization of quantum group generators.

\section{Final remarks}

In this article we compare two ways of force term generating in the model of nonrelativistic particle moving in
constant external field force. First of them uses the transformation
rules of classical space (\ref{trans1}) while the second one is based on the space-time noncommutativity (\ref{spaces}).  Particulary, we find for which functions $f(t)$ and $a_i(t)$ the generated force terms
are the same in both treatments. Finally, it should be noted that performed in this article considerations concern only the simplest model of nonrelativistic particle moving in constat force $\vec{F}$. However, they can be extended to the arbitrary norelativistic system and then, the necessary calculations become much more complicated but the general mechanism remains the same.

\section*{Acknowledgments}
The author would like to thank J. Lukierski
for valuable discussions. This paper has been financially  supported  by Polish
NCN grant No 2011/01/B/ST2/03354.

\end{document}